# Passivation-sensitive exciton finestructure produces excess Stokes shifts in colloidal quantum dots

O. Voznyy,[1] F. Fan,[1] A. Ip,[1] A. Kiani,[1] S. M. Thon,[2] K. Kemp,[1] L. Levina,[1] E. H. Sargent[1]*

[1] Department of Electrical and Computer Engineering, University of Toronto, 10 King's College Road, Toronto, Ontario, M5S 3G4, Canada

[2] Department of Electrical and Computer Engineering, Johns Hopkins University, 3400 North Charles Street, Baltimore, MD 21218-2608, USA

* Corresponding author: ted.sargent@utoronto.ca

**The excitonic finestructure of colloidal quantum dots (CQDs) is comprised of a manifold of transitions, of which only the lowest are populated and contribute to photoluminescence. This leads to a Stokes shift in emission relative to absorption. Here we show experimentally that the Stokes shift in Pb and Cd-based chalcogenide CQDs is correlated with the degree of surface passivation, and develop a model that explains how coupling to the surface affects the core electronic states. Dark and bright transitions can reorder and split, increasing the Stokes shift even without the formation of deep traps. Our findings resolve the highly-debated topic of excess Stokes shifts in PbS nanocrystals as due to parity-forbidden transitions instead of traps. We predict that the Stokes shift in PbS can be eliminated via core stoichiometry control, a critical step towards enhancing the open circuit voltage in quantum dot solar cells.**

The origin of the Stokes shift in colloidal quantum dots (CQDs) is closely related to the excitonic finestructure, which consists of a manifold of closely spaced optical transitions arising from the quasi-degenerate bandedge electronic states [1,2]. More than half of these transitions are forbidden by parity and spin selection rules and will be optically passive (dark excitons) and, thus, undetectable in absorption spectra [1–5]. However, photoluminescence from nanocrystals arises from the thermally-populated lowest levels. [3] The resulting shift between absorption and emission spectra is the Stokes shift, analogous to the behavior of molecular emitters, albeit mechanistically-distinct.

Controlling the Stokes shift in CQDs is of significant applied importance. Light-emitting diodes benefit from a bright ground state with short radiative lifetime that competes effectively with non-radiative processes. In contrast, in photoluminescence concentrators and lasers, a Stokes shift large enough to eliminate light reabsorption is desired. In photovoltaics, the difference between optical and electronic bandgaps leads to losses in the open-circuit voltage ($V_{OC}$) and should desirably be minimized. Lead sulfide is the material that has shown the greatest performance progress as a basis for quantum dot solar cells, yet it is particularly prone to a large Stokes shift whose origin remains unresolved [6,7].

Semi-empirical theoretical models based on effective mass [3,8,9], tight-binding [2,4,10] and pseudopotential [1,11] approximations all agree on the excitonic finestructure of CdSe and PbS CQDs. The Stokes shift is predicted to decrease monotonically with size [12]; however, the exact value of the Stokes shift is typically underestimated relative to what is observed experimentally [9]. In PbS, even atomically-resolved computational methods predict a zero Stokes shift (before the inclusion of Coulomb scattering effects) irrespective of nanocrystal shape and symmetry [1,10,11], in stark contrast with experimental observations of a very large shift reaching 180 meV for ~3 nm diameter PbS CQDs. Estimations of the Franck-Condon shift [13] and exchange interaction [1,9] predict at most a 20 meV dark-bright level splitting. Transient absorption and temperature-dependent photoluminescence measurements agree on the presence of a strongly split-off dark state [6,7]. This has been posited to have a surface-related origin.

In this letter, we develop a realistic *ab initio* model of nanocrystal surfaces to demonstrate that the Stokes shift arises from parity-forbidden transitions, and *not* from surface-related traps. Passivation affects the confinement and, consequently, the energy of the dark and bright states to a different degree, affecting the ordering of levels in the finestructure. We further explore the mechanisms that prevent the desired complete surface coverage with ligands, and propose several directions to improve the passivation of PbS CQDs and eliminate the Stokes shift entirely.

Calculations were performed using the SIESTA [14] and CP2K [15] codes. SIESTA allows for synthetic atoms with partial charges [16], a feature needed to investigate cases involving ideal surface passivation. The optical properties are assessed via the imaginary part of the dielectric function, and this includes the joint density of states altered by the dipole oscillator strength. Calculations were performed on 3 nm PbS and 2.5 nm CdSe CQDs prepared following published procedures [17,18]. A sphere is carved from a bulk crystal, and all singly-bonded atoms are discarded. We use cation-terminated CQDs, saturating the surface with Cl or oleic acid as representative ligands. The number of ligands is controlled to maintain charge neutrality and ensure trap-free nanocrystals [18,19]. Full input files and geometries are provided in Supplemental Material [20].

We carried out an initial suite of experimental studies to see if we could modulate the Stokes shift chemically. PbS CQDs were synthesized according to published protocols based on lead oxide and bis(trimethylsilyl)sulfide precursors and employing oleic acid as the ligand [21]. A large lead-to-sulfur precursor ratio of 2.7 was used; this has been shown to improve the absorption linewidth [12]. To investigate the effect of improved surface passivation, we added halide ligands via the introduction of tetrabutylammonium chloride (TBAC) salt [22,23]. TBAC in oleylamine solution, in amounts ranging from 0.18 to 0.6 mM, was injected at 60°C during the cooling stage of the synthesis. The total quantity of Cl incorporated onto the nanoparticle surface was quantified using X-ray photoelectron spectroscopy for different molar concentrations of TBAC [20]. For CdSe, a synthesis based on Cd oxide, elemental sulfur and oleic acid [24] was used without modification. CQDs were grown sufficiently small (~3 nm diameter for PbS, absorption edge of 950 nm, and ~2.5 nm diameter for CdSe, absorption edge of ~500 nm) to enhance the observed Stokes shift.

Post-synthesis washing or ligand exchanges led to an increase or decrease of the Stokes shift on nominally identical CQD sizes (Fig.1ab), suggesting that the shift is related to surface passivation. For a more controllable modification of the CQD surface, we explored the passivation of PbS CQDs using chloride ions. Incorporation of Cl has been demonstrated to reduce trap densities in CQD photovoltaic devices [22]. The amount of chloride on the surface was found to correlate well with the molar amount of TBAC used for treatment [20]. Importantly, we find that the amount of oleic acid was not affected, indicating that chloride does not displace the oleate but rather binds to otherwise unpassivated trenches not accessed by the bulkier organic ligand, improving the overall surface passivation [22].

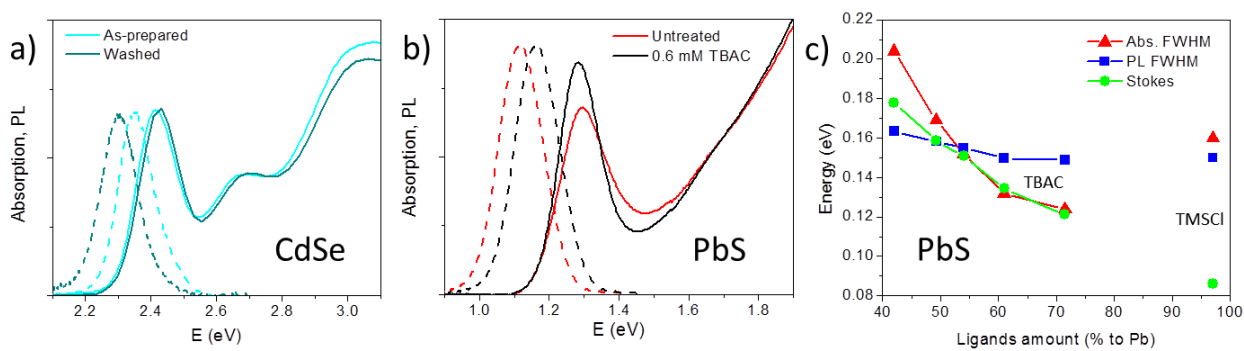

FIG. 1 (color online). Experimental optical absorption (solid lines) and emission (dashed lines) of CQD solutions. (a) Fresh and aged CdSe samples (light and dark blue lines, respectively). (b) PbS samples with and without chloride treatment (black and red lines, respectively). (c) Passivation-dependent linewidths and Stokes shifts for PbS CQDs with different amounts of incorporated chlorides. The largest ligand amount point was obtained using a trimethylsilyl chloride (TMSCl) ligand exchange (see text for details). The procedure for quantifying the surface ligand coverage and its relation to the amount of Cl introduced during the synthesis is described in Supplemental Material [20].

The addition of Cl passivation is accompanied by a decrease in the Stokes shift, which we studied as a function of a wide range of concentrations (Fig. 1c). The Stokes shift in PbS CQDs is reduced from 180 meV to 120 meV as the amount of chloride is increased. The smallest Stokes shift correlates with the narrowest exciton and higher energy transition peaks seen in the absorption spectra [20]. The same behavior was previously assigned to changes in the inhomogeneity of the size distribution [9]. Here we check this alternative hypothesis by monitoring the photoluminescence linewidth: we find that it is unaffected (Fig.1c), disconfirming the possibility of a major role for inhomogeneous broadening. In addition, certain samples showed increased Stokes shift without changes to their absorption peak width (Fig.1a) or bandtail position (Fig.1b).

Photoluminescence linewidth measurements also suggest that the emission is not trap-related, since trap emission is known to produce much broader (up to 350 meV) linewidth [25,26]. Unusually broad bandedge emission in PbSe and PbS CQDs is well documented [27] and can be assigned to phonon coupling [28] and multiplicity of the bandedge transitions participating in photoluminescence. The observation of photoluminescence linewidths that are broader than absorption linewidths indicates the presence of additional optically forbidden low-energy transitions that cannot participate in absorption but are participating in photoluminescence.

To clarify the origin of the emitting state in PbS CQDs, we have performed DFT simulations on nominally trap-free CQDs. In contrast to published semi-empirical models [1,4,10,11] and our own simulations with passivation by artificial atoms with non-integer charges (Fig.2a), we *do* observe a dark lowest-energy transition when *realistic* ligands (Cl, thiols or oleic acid) are used (Fig. 2b). Careful investigation of the electronic states confirms the expected 4-level manifolds of the 1S states on both the valence and conduction band sides. Previous models reported a more than 100 meV splitting of these manifolds due to intervalley coupling and asymmetry of the transverse and longitudinal effective masses in the L-valleys [4,10,11]; however, the lowest-energy transition was always reported to be bright even though dark states were found higher in energy. In the present case, the use of a realistic surface model causes these dark and bright transitions to reorder, leading to a large Stokes shift (Fig.2).

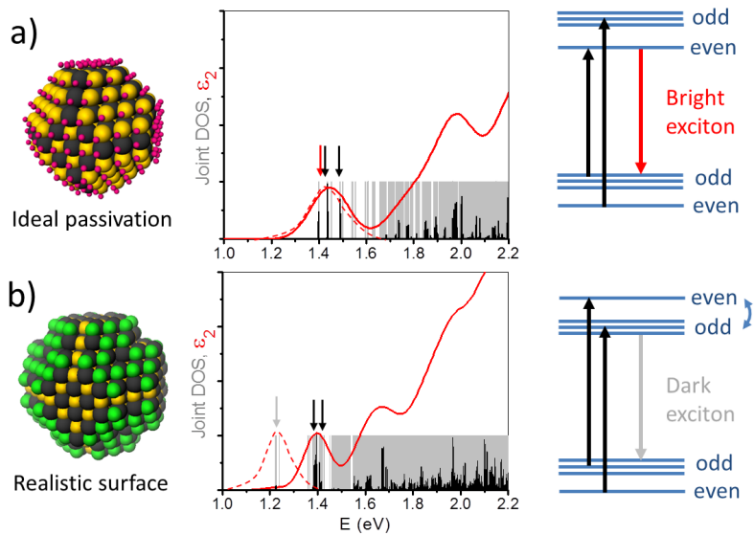

FIG. 2 (color online). Calculated optical spectra of 3 nm PbS CQDs and reordering of dark and bright transitions responsible for the Stokes shift. (a) Ideal passivation using artificial pseudo-hydrogen atoms with partial charge. (b) Passivation using realistic surface coverage by Cl ligands.

The effects of the surface on the oscillator strength and lifetime of the bandedge [29] and higher-lying transitions [5,30], as well as on exciton relaxation dynamics and the intensity of trap emission [31–34] have been reported. Our findings expand the spectrum of the surface effects to include reordering of the core bandedge states and Stokes shift magnitude.

To investigate the physical mechanisms by which surface passivation affects the core states, we evaluated the wavefunctions of the bandedge states in PbS and CdSe nanocrystals (Fig.3). We prevented the formation of localized trap states by eliminating surface atoms having more than two (out of six) dangling bonds in the case of PbS; and having more than one (out of four) dangling bond in the case of CdSe [20].

In PbS, core states of different parity within the four-fold degenerate $1S_e$ manifold are localized on different sublattices (Fig.3a). One of these sublattices encompasses the surface atoms and, as a result, is very sensitive to surface modifications. The other lattice does not include the surface atoms (but instead only the subsurface ones), and therefore is considerably less sensitive to surface effects such as passivation.

As a result of this effect, unpassivated dangling bonds on the surface cause the odd state to expand into the surface region, which in turn affects its effective size and thus its quantum-tuned energy levels. The even state is essentially unaffected. This discrimination as a function of the extent of surface-sampling results in surface-induced reordering of the electronic states. Saturating the surface with a higher amount of ligands, permitted by virtue of the small size of Cl ions, allows passivation of more dangling bonds and blocks the spillage of the wavefunction into the surface region (Fig.3). Similar behavior is found in CdSe, where severe underpassivation can also reorder odd and even states within the valence band manifold (Fig.3b). The reordering of *s* and *p* states has been invoked previously to explain the long radiative lifetime in small CdS CQDs, but the effect was found to be due to the intrinsic electronic structure of CdS rather than a surface effect [35].

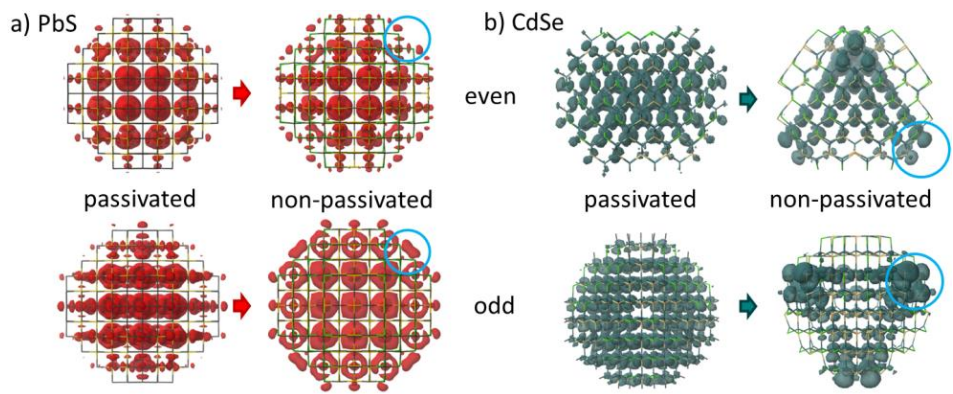

FIG. 3 (color online). Surface-dependent symmetry and quantum confinement of CQD core electronic states. (a) PbS conduction band states, (b) CdSe valence band states. Blue circles highlight the regions where the wavefunction spills out into the dangling bonds region.

A difference in spatial extent of the electron and hole wavefunctions can lead to a weak dipole moment when an exciton is formed inside the CQD. This produces the potential for polarization when a solvent with a high dielectric constant is used. We tested the system for solvatochromism by measuring the photoluminescence lifetime of CQDs dispersed in hexane (dielectric constant of 1.88) and chloroform (dielectric constant of 4.81). We observed a weak lifetime change from 1.1 μs to 1.4 μs, consistent with the presence of a dipole and the possibility of controlling the electron-hole overlap by changing only one of the wavefunctions. The effect is, however, weaker than expected for a strongly localized (trap) state.

We further investigated whether the Stokes shift could, perhaps with the aid of future novel experimental strategies, be dramatically reduced or eliminated despite the surface-induced effects due to realistic ligands. First we attempted to build CQD models exhibiting higher ligand coverage. We find, however, that the number of anionic ligands required to cover all dangling bonds fails to satisfy the charge neutrality condition [18,25]. For a given nanocrystal with the predefined cation-to-anion stoichiometry, any excess ligand results in either charging or electronic doping [18]. In the case of the 3 nm spherical PbS CQD in Fig.2b the required excess doping is quite substantial, amounting to 42 electrons. This significantly destabilizes the nanocrystal, and balance is restored by desorbing excess ligands. For example, we calculated the barrier for desorption of Cl ligands as $Cl_2$ molecules to be ~5 eV for a charge-balanced nanocrystal and found that this figure diminishes to 1.7 eV when the CQD is doped to excess. Clearly, the resulting nanocrystal exposes unpassivated dangling bonds and is prone to a large Stokes shift.

In light of this finding, we sought methods that would retain excess anionic ligands and achieve a complete passivation of cation surface sites. This would require changing the cation:anion stoichiometry in the core. We explored two methods to achieve this: changing the CQD shape, and replacing the surface-exposed anions with lower-valence atomic ligands (Fig.4). We note that simply charging or doping the original structure in Fig.2b to saturate more Pb sites with ligands does not solve the problem because surface-exposed sulfur sites remain unpassivated.

We found that the addition of extra $Pb^{2+}$ cations to the ridges of a spherical PbS CQD allows the addition of twice the number of ligands having charge 1- (Fig.4a). This can eventually saturate all surface Pb sites and, at the same time, eliminate exposed S sites. This brightens the lowest-energy optical transition and also distinctly changes the absorption spectrum, which can be used as a signature to look for in future studies.

We then turned to a different strategy: replace the $S^{2-}$ core anions with a $halide^{1-}$ (Fig.4b) to balance charge without affecting the CQD structural integrity. The resulting CQD indeed exhibits no low-energy dark states, completely eliminating the Stokes shift.

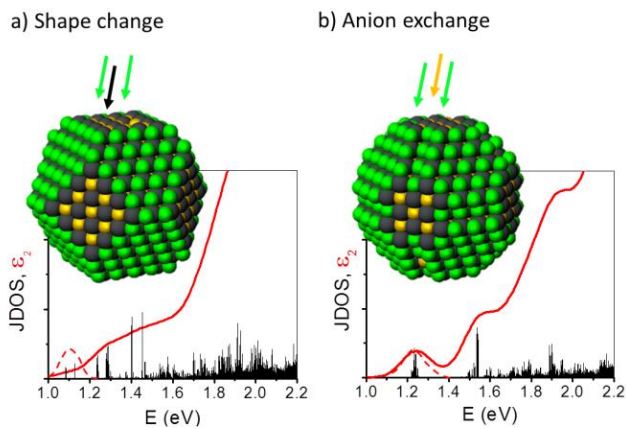

FIG. 4 (color online). Improving the passivation to achieve lower Stokes shift. (a) Addition of extra cations to counterbalance excess anionic ligands. (b) Incorporation of halides into the sulfur lattice.

Inspired by these findings we sought to increase the saturation of CQD surfaces with ligands experimentally. We used a recently reported approach based on a complete ligand exchange from oleic acid to chloride (using trimethylsilyl chloride as a source) and amine ligands [36,37]. We measured a further reduction in the Stokes shift down to 86 meV (right-most point in Fig.1c) which correlates well with the increase in ligand coverage [20]. It should be noted that this synthesis technique results in significantly broader absorption linewidth, indicating that it is not a prerequisite for a smaller Stokes shift.

Complete elimination of a Stokes shift may require advances in inorganic synthetic chemistry aimed at more complete replacement of surface sulfur. Indeed hints in this direction have very recently been observed in studies in which molecular $I_2$ displaces sulfur in the PbS lattice more efficiently [38–40].

In conclusion, we have found a link between surface passivation of nanocrystals and their excitonic finestructure. This allowed us to explain the dark ground state in PbS nanocrystals with reference to parity-forbidden transitions between core states. The explanation involves a role for surface passivation, but is distinct from the more familiar idea of surface traps. Experiments confirm a fully two-fold reduction of the Stokes shift with improved surface passivation. Theory predicts that further progress towards the ultimate goal of eliminating the dark ground state splitting may benefit from further advances in controlling nanocrystal shape and stoichiometry.

We thank Grant Walters and James Fan for help with photoluminescence measurements. This publication is based in part on work supported by Award KUS-11-009-21, made by King Abdullah University of Science and Technology (KAUST), by the Ontario Research Fund Research Excellence Program, and by the Natural Sciences and Engineering Research Council (NSERC) of Canada. Computations were performed on the GPC supercomputer at the SciNet

HPC Consortium [41]. SciNet is funded by: the Canada Foundation for Innovation under the auspices of Compute Canada; the Government of Ontario; Ontario Research Fund - Research Excellence; and the University of Toronto.